# The impact of process steps on nearly ideal subthreshold slope in 300-mm compatible InGaZnO TFTs

Hongwei Tang, Dennis Lin, Subhali Subhechha, Adrian Chasin, Daisuke Matsubayashi, Michiel van Setten, Yiqun Wan, Harold Dekkers, Jie Li, Shruthi Subramanian, Zhuo Chen, Nouredine Rassoul, Yuchao Jiang, Jan Van Houdt, Valeri Afanas'ev, and Attilio Belmonte

*Abstract*— **While we demonstrate a back-gated (BG) amorphous Indium-Gallium-Zinc-Oxide (a-IGZO) transistors with a nearly ideal subthreshold slope (SS) ~ 60 mV/dec. However, SS degrades when a top-gated (TG) configuration is implemented. The energy distribution of traps inferred from temperature-dependent (T = 4 K - 300 K) and multi-frequency (f = 1 kHz - 100 kHz) admittance measurements, reveals a much higher trap density in TG devices. By analyzing the impact of each process step and conducting forming gas anneal (FGA) experiments, we reveal the role of hydrogen in the deterioration of the SS in the IGZO-based transistors.**

## I. INTRODUCTION

The ultra-low off-current ($I_{off}$ <$10^{-20}$ A/μm) [1] and back-of-end-line (BEOL) process compatibility [2] of amorphous Indium-Gallium-Zinc-Oxide (a-IGZO) has made IGZO transistors highly attractive for a plethora of applications, ranging from display technology [3], [4] to capacitorless 2T0C DRAM [1], [5], [6], $4F^2$ DRAM [7], compute-in-memory [8], [9], and monolithic 3D integration [10]. In memory applications, the retention performance of DRAM cells is mainly determined by the $I_{off}$ of write transistors utilizing IGZO channels [5], [6], [7], which is intrinsically linked to the subthreshold slope (SS) [11].

Previous reports have demonstrated that SS in IGZO thin-film-transistors (TFTs) can be affected by many factors, such as material's quality, processes, and device structures [12], [13]. However, the fundamental mechanisms determining SS in IGZO devices have not been fully investigated. SS is typically attributed to interface traps, but the common method to correlate SS degradation with interface trap density ($D_{it}$) through the relation SS $\approx k_BT/q \times \ln(10) \times (1+qD_{it}/C_{ox})$ [14], provides limited information on the energy distribution of traps, as the $D_{it}$ represents an integrated result over the sub-gap energy range in which thermal equilibrium trap occupancy is maintained.

Previously, we demonstrated an experimental method to extract the energy profile of the sub-gap density of states (DoS), corresponding to the total trap density per energy interval ($D_t$), in IGZO TFTs [16]. The shortest channel length ($L_{ch}$) in our prior experiments was limited to 3 μm, which was insufficient to eliminate the effects of parasitic resistances.

In this work, we apply this methodology to scaled IGZO transistors with $L_{ch}$ < 1 μm fabricated on 300-mm wafers for the first time, enabling a reliable $D_t$ extraction in back-gated (BG) and top-gated (TG) devices. We elucidate the mechanisms causing the SS degradation when transitioning from BG to TG structures, and we reveal the role of hydrogen by correlating the forming gas anneals (FGA) and sub-gap DoS extraction.

## II. DEVICE FABRICATION

*Back-gated structure* – The device stack and process flow are illustrated in **Fig. 1(a)** and **1(b)**. On a 300-mm highly doped p-Si substrate, a 5-nm $Al_2O_3$ layer is deposited as bottom gate dielectric by atomic layer deposition (ALD). Then, a 12-nm a-IGZO film is deposited by physical-vapor deposition (PVD), followed by active patterning. Contact trenches are then formed and filled with TiN and W metals, followed by an additional metallization layer.

*Top-gated structure* – The device stack and TEM cross-section image are shown in **Fig. 1(c)** and **1(d)**. The TG architecture is optimized with an oxygen tunnel layer and raised contacts using c-axis-aligned-crystalline (CAAC) IGZO layer, to maximize the on current performance [5], [15]. More details about the TG processing flow are reported in [16].

Temperature-dependent electrical characterization is performed using a helium-cooled vacuum probe station. The capacitance/conductance is measured by Agilent E4980A LCR meter with a parallel $C_p$-G mode. The $I_d$-$V_g$ characterization is

$^{1}$This paragraph of the first footnote will contain the date on which you submitted your paper for review. The authors would like to thank the support of imec's Industrial Partners in the Active Memory Program. This work has been enabled in part by the NanoIC pilot line. The acquisition and operation are jointly funded by the Chips Joint Undertaking, through the European Union's Digital Europe (101183266) and Horizon Europe programs (101183277), as well as by the participating states Belgium (Flanders), France, Germany, Finland, Ireland and Romania. For more information, visit nanoic-project.eu.

Hongwei Tang, Zhuo Chen, Jan Van Houdt and Valeri Afanas'ev are with the Department of Physics and Astronomy, Katholieke Universiteit Leuven,3001 Leuven, Belgium, and also with IMEC, 3001 Leuven, Belgium (e-mail: Hongwei.Tang@imec.be).

Dennis Lin, Subhali Subhechha, Adrian Chasin, Daisuke Matsubayashi, Michiel van Setten, Yiqun Wan, Harold Dekkers, Jie Li, Shruthi Subramanian, Nouredine Rassoul, Yuchao Jiang, Gouri Sankar Kar, and Attilio Belmonte are with IMEC, 3001 Leuven, Belgium (e-mail: Attilio.Belmonte@imec.be).

carried out with an Agilent B1500 parameter analyzer. For BG devices, the gate voltage is swept on the heavily p-doped silicon substrate.

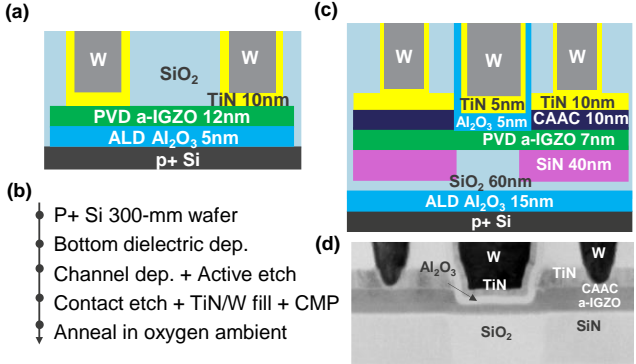

Fig. 1. (a) Structure and (b) process flow of back-gated (BG) a-IGZO devices; (c) Structure and (d) corresponding TEM image of top-gated (TG) a-IGZO devices. Devices are fabricated on 300-mm wafers.

## III. RESULTS AND DISCUSSION

### A. Experimental $I_d$-$V_g$ and SS behavior

The $I_d$-$V_g$ characteristics of BG and TG a-IGZO transistors (including 70 devices) at room temperature are shown in **Fig. 2(a)**. Comparing the SS for BG and TG devices in **Fig. 2(b)**, we observe that the BG devices systematically show steeper SS, with lowest values $\sim 60$ mV/dec. Considering that the increase in SS is determined by electron traps in the bandgap, characterization of the sub-gap trap DoS should provide insight for understanding this SS difference.

### B. Sub-gap trap DoS characterization

The *ac* conductance method is a well-known technique for extracting sub-gap DoS in semiconductor MOS devices from their capacitance-voltage (C-V) and conductance-voltage (G-V) properties. Here we used a multi-finger transistor [17] for C-V/G-V measurements, by maximizing the channel area and ensuring channel depletion while keeping the $L_{ch}$ short (< 1 μm). The measured C-V under different temperatures of BG devices with $L_{ch}$ = 330 nm are plotted in **Fig. 3(a)**. No frequency-dispersion is observed in the C-V characteristics for $L_{ch}$ = 330 nm at T = 300 K, indicating negligible trap response. At temperatures below 100 K, the frequency-dispersion begins to appear.

The C-V characteristics of the TG device with $L_{ch}$ = 300 nm are shown in **Fig. 3(b)**. Unlike the BG devices, and consistent with the SS degradation, the frequency-dispersion is clearly observed in the low-C region of the C-V curve already at T = 300 K. The 'triangular' shape is the signature of the trap response [18]. As the temperature decreases to 120 K, the frequency dispersion becomes more pronounced (the inset in **Fig. 3(b)**), indicating a response of traps closer to conduction band edge ($E_c$).

The parallel conductance $G_p/Aq\omega$ is calculated as: $G_p/Aq\omega = \omega C_{ox}^2 G_m/(G_m^2 + \omega^2(C_{ox}-C_m)^2)$ [19], where $C_m$ and $G_m$ are measured capacitance and conductance, $C_{ox}$ is dielectric capacitance, $A$ is channel area, $q$ is electron charge and $\omega = 2\pi f$ is angular frequency. The parabolic $G_p/Aq\omega$ curves of BG devices under T = 4 ~ 100 K are shown in **Fig. 3(c)**. From that, $D_t$ can be extracted as $D_t \approx f(\sigma_s) \times (G_p/Aq\omega)_{peak}$, where $f(\sigma_s)$ is a pre-factor ranging from 2.5 to 4, depending on the width of each parabolic curve [19]. The trap energy ($E_c-E_t$) is estimated from the frequency of the peak position ($f_{peak}$) using Shockley-Read-Hall model [12] as $E_c-E_t = k_B T \ln(\sigma_n v_{th} N_c \tau_n)$, where $\sigma_n$ is electron capture cross-section ($\approx 10^{-14}$ cm²), $v_{th}$ is the thermal velocity, $N_C$ is effective density of states in the conduction band, and $\tau_n$ is trap time constant ($\approx 2/(2\pi f_{peak})$) [19].

**Fig. 3(d)** compares the sub-gap DoS results extracted from BG and TG devices, along with data from literature [21], [22]. Both devices reveal a DoS tail stretching from the conduction band edge. At T = 300 K, no data points for BG devices are available within the accessible energy range because $D_t$ is below the detection limit. In comparison, BG devices show a lower DoS than TG structure, which is consistent with the observed SS performances.

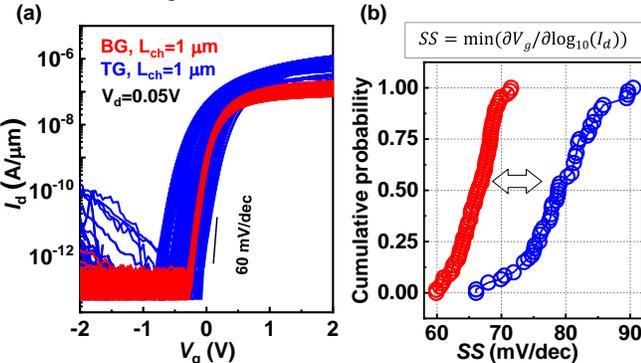

Fig. 2. (a) Measured $I_d$-$V_g$ characteristics and (b) extracted SS of BG and TG a-IGZO transistors (70 devices tested across the wafer) with channel width ($W_{ch}$) of 1 μm.

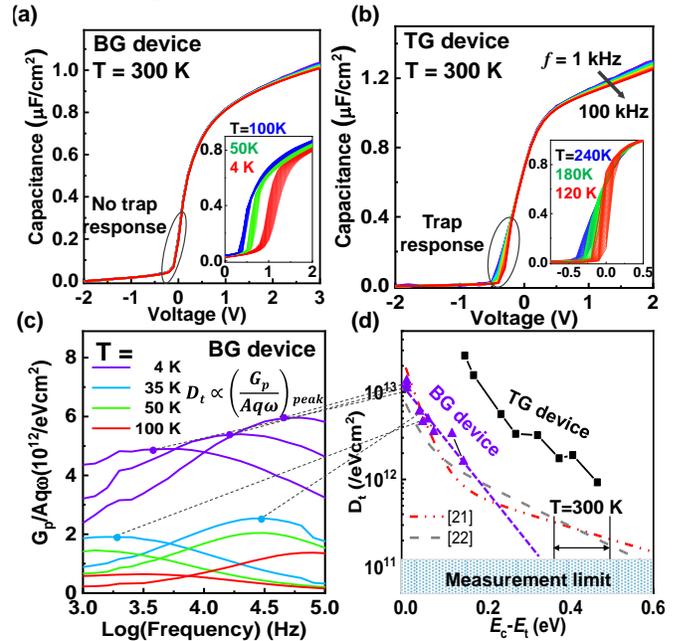

Fig. 3. Measured C-V at different temperatures of (a) BG IGZO devices and (b) TG devices. The measurement frequency ranges from 1 kHz to 100 kHz, covering 21 frequency conditions. The insets present the C-V data for different temperatures, with each temperature plotted in a single color. (c) Calculated $G_p/Aq\omega$ under different temperatures for BG devices. (d) Extracted $D_t$ from the $G_p/Aq\omega$ plot. Each $G_p/Aq\omega$ peak point

corresponds to a $D_t$ point. We highlighted the detectable trap energy range at T = 300 K, where $D_t$ of BG devices are below measurement limit.

## C. Mechanisms of SS degradation

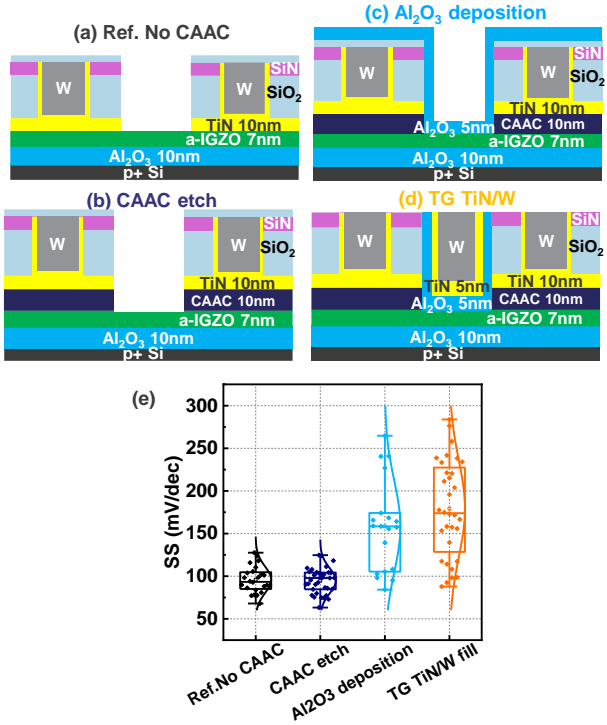

Fig. 4. (a)-(d) Four wafers with different stacks to isolate each additional process step in the transition from BG to TG. An anneal in $O_2$ ambient (250 °C/15 min) is applied to all wafers to obtain $V_t$ > -1 V. (e) Box plot of extracted SS after each process step. All devices are tested by sweeping $V_g$ on the bottom substrate.

Several differences between BG and TG devices in **Fig.1**, including channel thicknesses, material stacks and processing steps, makes these structures unsuitable for detailed investigations into the mechanisms behind SS and $D_t$ variation. While our previous work shows IGZO channel thickness (5~10 nm) does not induce a mismatch in SS [23], we concentrate on process steps that are expected to have a strong impact [24], particularly those steps that expose the IGZO channel and have the potential to damage it leading to SS degradation.

To address this, we carefully designed a controlled experiment including four wafers (**Fig. 4(a)-4(d)**) to isolate the influence of specific process steps (and also device stack) while keeping all other process conditions unchanged. The SS results in **Fig. 4(e)** shows the etching of the CAAC layer has no noticeable impact. However, a strong degradation is observed after ALD deposition of $Al_2O_3$, with further deterioration after TiN/W gate metals deposition. For both process steps, hydrogen is abundant and may be responsible for the SS degradation. During the top dielectric deposition, hydrogen from the deposition precursor (TMA, Al $(CH_3)_3$) and $H_2O$ may introduce additional trap states at the interface or within the IGZO channel. Additionally, the interaction of the aluminum precursors with the IGZO layer may also cause extra traps due to oxygen scavenging effect [25]. On the other hand, we already reported the role of the chemical-vapor-deposited (CVD) W fill as hydrogen source in IGZO TFTs [24].

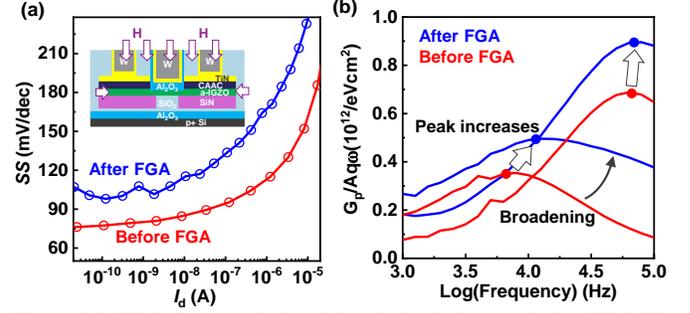

Fig. 5. (a) SS-$I_d$ characteristics of the device before and after FGA. The inset is the schematic of hydrogen diffusion process in a TG device. (b) $G_p/A q \omega$ of the TG device before and after FGA.

To examine the role of hydrogen, we conducted an FGA experiment (in 10% $H_2$/90% $N_2$ ambient at 400 °C for 30 minutes) on TG IGZO devices (the inset in **Fig. 5(a)**) with $L_{ch}$ = 300 nm, revealing a strong SS deterioration after the anneal (**Fig. 5(a)**). Similar degradation phenomena is observed in BG devices after FGA. The $G_p/A q \omega$ peaks (**Fig. 5(b)**), as extracted from measured admittance results of the TG device, become higher and broader, indicating higher $D_t$. These results prove that hydrogen introduces additional sub-gap trap DoS in IGZO devices and thus degrades SS.

Note that the microscopic nature of the new traps remains unclear, and it is uncertain whether these traps are directly related to hydrogen or result from hydrogen-induced changes in the IGZO film. To gain deeper insights, advanced material characterizations, such as secondary ion mass spectrometry (SIMS) could be utilized to investigate hydrogen-related bonds and concentrations before and after annealing. This remains an interesting topic for future exploration.

Nevertheless, our experiments provide a clear guideline for the fabrication of IGZO transistors: to preserve the nearly ideal SS observed in BG devices, reducing the H content during the process of TG devices is of utmost importance.

## IV. CONCLUSION

We demonstrated that back-gated IGZO-TFTs fabricated on 300-mm wafers can achieve nearly ideal SS ≈ 60 mV/dec. By developing an experimental methodology to extract the energy-dependent sub-gap DoS through admittance measurements, we revealed that device architecture and the process flow play a critical role in determining the SS. We identified the key process steps resulting in the SS degradation in top-gated devices compared to back-gated devices, unveiling the detrimental role of hydrogen during fabrication. The insight provided by this work is instrumental for future fabrication of scaled IGZO-TFTs with nearly ideal SS, targeting multiple low-power applications by enabling high $I_{ON}/I_{OFF}$ ratio in a reduced operating voltage range.